\documentclass[11pt,a4paper,english,oneside,parskip=false]{scrartcl} 
\usepackage[T1]{fontenc}
\usepackage[utf8]{inputenc}
\usepackage[english]{babel}

\usepackage{amsmath}
\usepackage{amsfonts}
\usepackage{amssymb}

\usepackage{mathpazo}
\usepackage{pifont}

\usepackage[backend=biber,style=authoryear-comp,maxcitenames=2,maxbibnames=6,minbibnames=6,sorting=nyt,uniquelist=false,uniquename=false]{biblatex}

\usepackage{hyphenat}
\usepackage{url}
\usepackage{hyperref}
\hypersetup{%
	final=true,%
	colorlinks=true,%
	linkcolor=black,%
	citecolor=black,%
	urlcolor=black,%
}
\usepackage[babel=true]{csquotes}
\usepackage{graphicx}
\usepackage{caption}
\usepackage{subcaption}
\usepackage{booktabs}

\usepackage{enumitem}
\newlist{compactitemize}{itemize}{3}
\setlist[compactitemize]{nosep,label={\raisebox{.1em}{\small\textbullet}}}
\newlist{compactenumerate}{enumerate}{3}
\setlist[compactenumerate]{nosep,label=\arabic*.}

\setkomafont{disposition}{\rmfamily}
\newcommand{\RR}{\mathbb{R}}
\newcommand{\tn}[1]{\textnormal{#1}}
\newcommand{\institute}[1]{\makebox[0pt][l]{$^{\text{#1}}$}\hspace{5pt}}

\addto\extrasenglish{%

}

\begin{document}

\title{\vspace*{-6pt} Modeling Polyp Activity of \textit{Paragorgia arborea} Using Supervised Learning}
\author{Arne Johanson,\institute{a} Sascha Flögel,\institute{b} Wolf\hyp{}Christian Dullo,\institute{b}\\Peter Linke,\institute{b} Wilhelm Hasselbring\institute{a}\\[-1pt] \footnotesize \institute{a}\hspace{-2pt} Software Engineering Group, Kiel University, Germany\\[-7pt]\footnotesize\institute{b}\hspace{-2pt} GEOMAR Helmholtz Centre for Ocean Research, Kiel, Germany}
\date{}

\maketitle

\vspace*{-7pt} 
\textbf{Abstract}---While the distribution patterns of cold\hyp{}water corals, such as \textit{Paragorgia arborea}, have received increasing attention in recent studies, little is known about their \textit{in situ} activity patterns. 
In this paper, we examine polyp activity in \textit{P. arborea} using machine learning techniques to analyze high\hyp{}resolution time series data and photographs obtained from an autonomous lander cluster deployed in the Stjernsund, Norway. 
An interactive illustration of the models derived in this paper is provided online as supplementary material. 

We find that the best predictor of the degree of extension of the coral polyps is current direction with a lag of three hours. 
Other variables that are not directly associated with water currents, such as temperature and salinity, offer much less information con\-cerning polyp activity. 
Interestingly, the degree of polyp extension can be predicted more reliably by sampling the laminar flows in the water column above the measurement site than by sampling the more turbulent flows in the direct vicinity of the corals. 

Our results show that the activity patterns of the \textit{P. arborea} polyps are governed by the strong tidal current regime of the Stjernsund. 
It appears that \textit{P. arborea} does not react to shorter changes in the ambient current regime but instead adjusts its behavior in accordance with the large\hyp{}scale pattern of the tidal cycle itself in order to optimize nutrient uptake. 

\section{Introduction}

Cold\hyp{}water corals (CWCs) such as \textit{Paragorgia arborea} and \textit{Lophelia pertusa} can be found on continental shelves, slopes, and seamounts all over the world. 
Like tropical coral reefs, which inhabit shallower and warmer waters, CWC reefs are associated with high biodiversity as they provide habitat to many other species \parencite{roberts2014,roberts2006}. 

The distribution of corals such as \textit{L. pertusa} and related fauna in the north\hyp{}east Atlantic is found to be linked to a particular density envelope ($\sigma_\theta$ between $27.35$ and $27.65$~kg~m$^{-3}$), which may control the dispersion of larvae and/or nutrient enrichment \parencite{dullo2008}.
Locally, CWCs occur at sites that offer hard surfaces for the corals to attach to and favorable current regimes that supply suspended food material to the filter feeders \parencite{davies2009}. 
Colonies of \textit{P. arborea} orient themselves perpendicular to the laminar water flow of tidal currents to maximize nutrient supply \parencite{mortensen2005}. 

While the study of distribution patterns of CWCs has received increasing attention during the past two decades \parencite[see, e.g.,][]{yesson2012,tittensor2009,davies2008predicting}, little is known about their \textit{in situ} activity patterns (i.e., the processes governing the extension and retraction of their polyps). 
We differentiate the two main stages of either polyp extension or retraction because these states represent different polyp activity regimes:  
while the polyps are extended, they capture food particles, and while they are retracted, they digest the previously captured food or are inactive \parencite{mortensen2005}. 
Retraction is potentially also associated with increased environmental stress such as, e.g., intensified currents or particle loads \parencite{larsson2011,larsson2013}. 
However, \textcite{purser2015} states that polyp \enquote{retraction as a response to environmental stress has been difficult to establish unambiguously.} 

To study polyp activity in \textit{P. arborea}, we analyzed high\hyp{}resolution time series data and photographs of polyp activity obtained from an autonomous lander cluster deployed in the Stjernsund, Norway in June 2012 at a water depth of 215 m. 
For this analysis, we employed machine learning techniques, specifically supervised learning \parencite{shalev2014}. 
An interactive illustration of the models derived in this paper can be explored online via OceanTEA \parencite{CI2016}.\footnote{Follow the link provided at: \url{https://github.com/a-johanson/paragorgia-arborea-activity}} 


\section{Materials and Methods} \label{sec:methods}

\subsection{Study Area and Instrumentation}

\begin{figure}
\centering
\includegraphics[width=\linewidth]{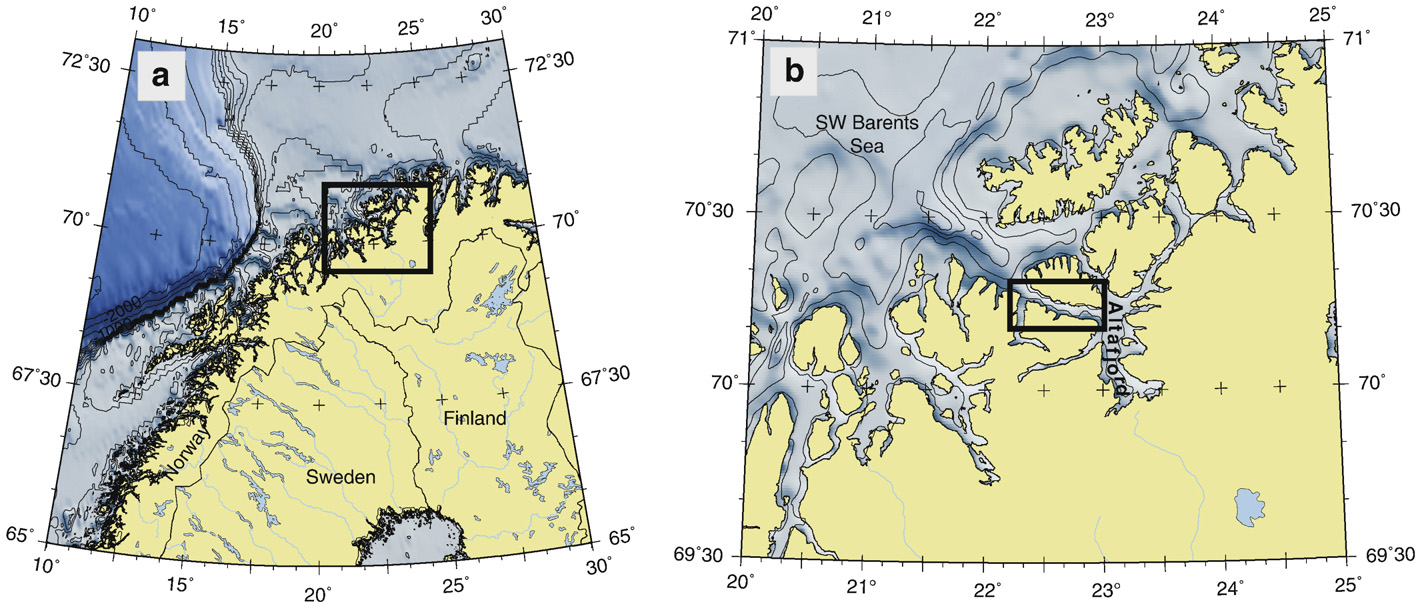}
\caption{a) The Finnmark district of northern Norway with b) the Stjernsund \parencite[from][]{rueggeberg2011}.}
\label{fig:map_region}
\end{figure}

In June 2012, the autonomous lander cluster MoLab (Modular Multidisciplinary Ocean Laboratory) was installed on a CWC reef in the Stjernsund, Norway for four months. 
The Stjernsund, which is a 30 km long and up to $3.5$ km wide sound connecting the North Atlantic with the Altafjord, is located at $70.5^\circ$N and $22.5^\circ$E (\autoref{fig:map_region}). 
The lander cluster was deployed at a water depth of about 200 to 350 meters on a morainic sill, which houses one of the northernmost CWC reefs in Europe \parencite{rueggeberg2011}. 

\begin{figure}
\centering
\includegraphics[width=\linewidth]{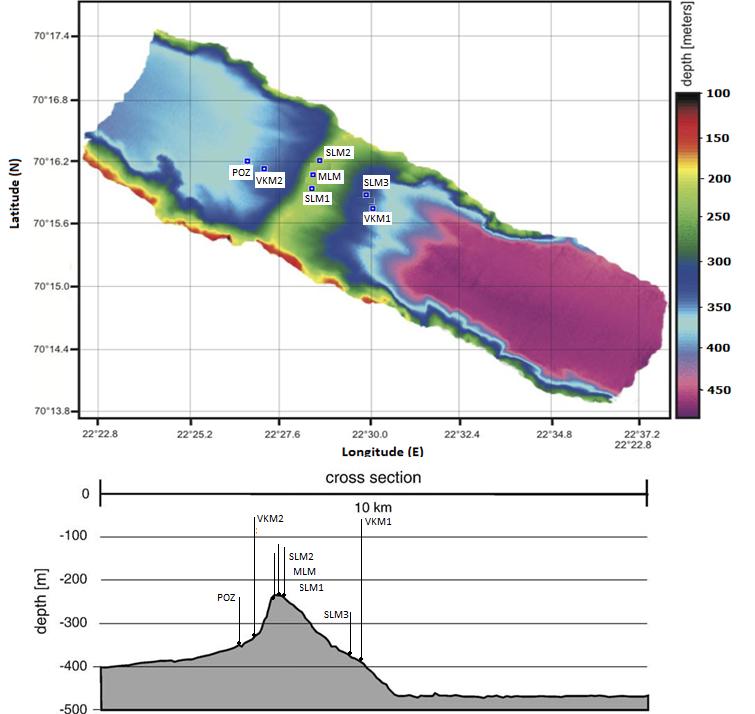}
\caption{Swath bathymetry (top) of the Stjernsund and sill cross section (bottom) with the locations of the MoLab systems \parencite[main lander (MLM), satellite landers (SLM, POZ) and moorings (VKM); from][]{ashastina2013}.}
\label{fig:map_landers}
\end{figure}

The MoLab configuration deployed in the Stjernsund consisted of five lander systems (one main lander and four satellite landers) and two moorings (\autoref{fig:map_landers}). 
Each lander system was equipped with sensors to measure temperature, conductivity, pressure, pH, fluorescence, turbidity, as well as flow direction and velocity in the water column using acoustic Doppler current profilers (ADCPs). 

In this study, we analyzed the time series data from the MoLab main lander MLM, which was equipped with
\begin{compactenumerate}
	\item a fluorescence and turbidity sensor (Wetlabs FLNTU (RT)),
	\item a pH sensor (Sea\hyp{}Bird Elecronics SBE 27),
	\item a temperature, conductivity, and (digiquartz) pressure sensor attached to an SBE 16plus V2 SeaCAT,
	\item an upward\hyp{}facing ADCP (RDI 300 KHz),
	\item a downward\hyp{}facing ADCP (RDI 1200 KHz),
	\item and a downward\hyp{}facing digital camera system to capture still images (12 megapixels).
\end{compactenumerate}
The instruments to measure pH, fluorescence, and turbidity on this lander did not function correctly. 
Therefore, we studied only the data from the temperature, conductivity, and pressure sensor as well as from the ADCPs, which all acquired measurements in regular 10 minute intervals. 

The ADCPs measured flow direction and velocity averaged over bins of $1$ m (up) and $0.1$ m (down). 
The height of the first bin was $3.22$ m (up) and $0.41$ m (down). 
We ignored the first three bins of the upward\hyp{}facing ADCP and the first bin of the downward\hyp{}facing instrument because of excessive noise in these bins. 
For the upward\hyp{}facing ADCP, we included $46$ bins ($5.22$ to $51.22$ m above the lander system) in our analysis, and for the downward\hyp{}facing ADCP $11$ bins ($0.41$ to $1.51$ m below the lander system).

\subsection{Photographs of Polyp Activity} \label{subsec:photographs}

\begin{figure}
\centering
\includegraphics[width=0.9\linewidth]{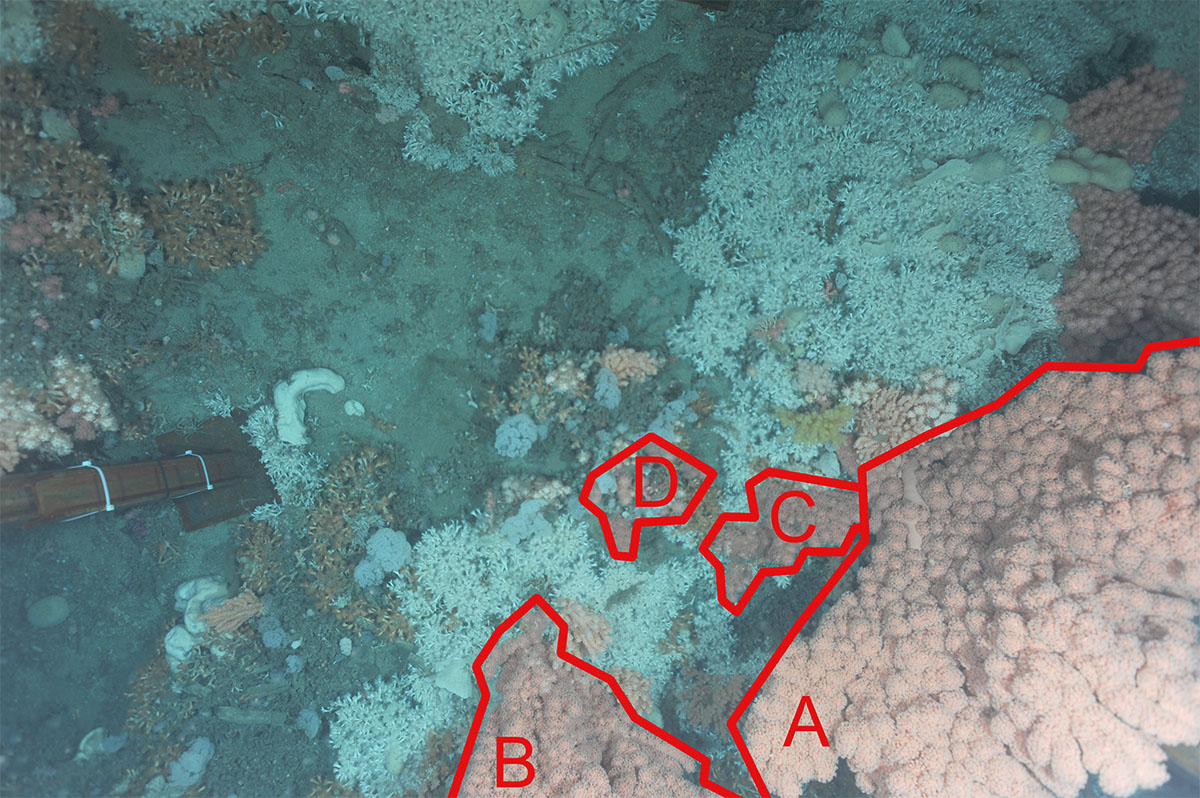}
\caption{Full camera frame with \textit{P. arborea} groups A, B, C, and D.}
\label{fig:full_scene_label}
\end{figure}

The downward\hyp{}facing camera of the main lander captured $258$ images in total, one image every hour (from 3:00 on June 5, 2012 to 13:00 on June 16, 2012 (UTC); at a few sample points, the camera failed to record an image).  
An example of such a photograph is displayed in \autoref{fig:full_scene_label}. 
The lower right part of the image shows multiple \textit{P. arborea} colonies, which we divided into four groups (A, B, C, and D). 

The time interval between two photographs is short enough to capture the relatively stable behavioral cycle of the coral polyps. 
Otherwise, as discussed further below in \autoref{subsec:large_vs_fine_scale}, it would be very unlikely to observe the regular cycles of extended and retracted states present in our dataset. 

We assigned a label to each frame captured by the camera describing whether the polyps of the \textit{P. arborea} corals are \emph{extended} or \emph{retracted} in this frame. 
For labeling the images, we considered only the coral groups A--D and ignored all other \textit{P. arborea} colonies (they cannot be classified reliably in some images because they are out of focus and/or underexposed). 

\begin{figure}
\centering
\begin{subfigure}[b]{0.68\linewidth}
    \includegraphics[width=\textwidth]{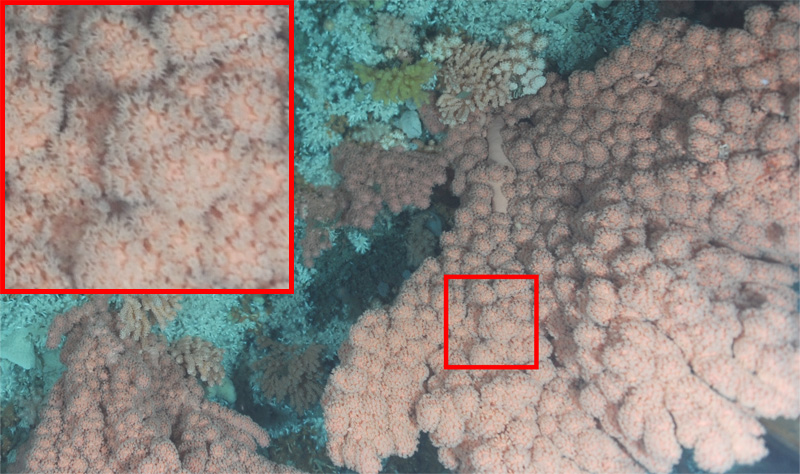}
    \caption{Extended\vspace*{6pt}}
    \label{fig:label_extended}
\end{subfigure}
\begin{subfigure}[b]{0.68\linewidth}
    \includegraphics[width=\textwidth]{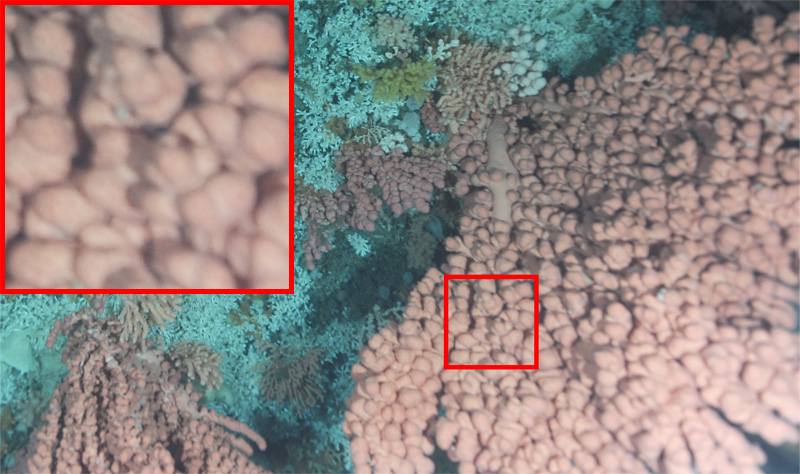}
    \caption{Retracted\vspace*{6pt}}
    \label{fig:label_retracted}
\end{subfigure}
\begin{subfigure}[b]{0.68\linewidth}
    \includegraphics[width=\textwidth]{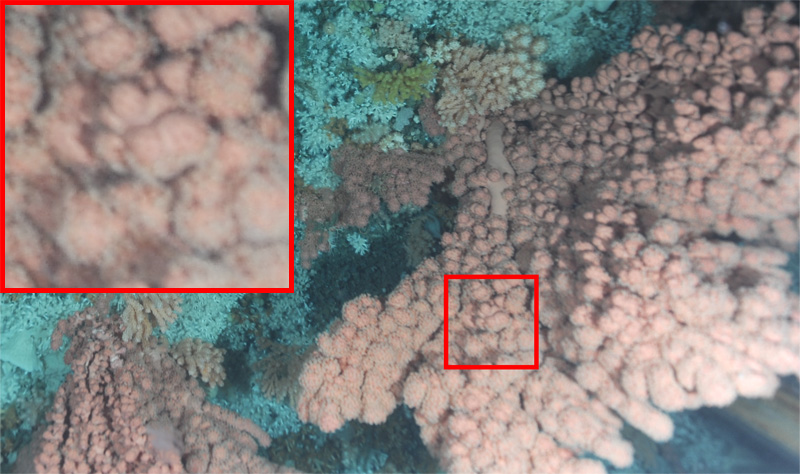}
    \caption{Intermediate}
    \label{fig:label_interm}
\end{subfigure}
\caption{Coral group A in the \emph{extended}, \emph{retracted}, and \emph{intermediate} state with full\hyp{}resolution inlets.}
\label{fig:label_examples}
\end{figure}

Before labeling the frames, we first preprocessed the raw images using Imagemagick\footnote{\url{http://www.imagemagick.org}} to crop each image to its lower right corner and to apply automatic gamma and color level corrections. 
Then, for image $i$, we manually assigned a value to each of the coral groups A--D according to the following decision rules (exemplary for group A):
\begin{itemize}
	\item $A_i = 1$ if group A is fully or almost fully ($>90 \%$ of area) covered with fuzzy polyps in image $i$ (for an example, see \autoref{fig:label_extended}).
	\item $A_i = 0$ if no or almost no ($<10\%$) fuzzy polyps are visible (see \autoref{fig:label_retracted}).
	\item $A_i = 0.5$ in all other cases (see, e.g., \autoref{fig:label_interm}).
\end{itemize}
We determined the degree of extension of the polyps $\eta_i$ in image $i$ as a linear combination of the degree of extension of the individual groups:
\begin{equation} \label{eq:linear_combination_observations}
	\eta_i = 0.7\cdot A_i + 0.2\cdot B_i + 0.07\cdot C_i + 0.3\cdot D_i
\end{equation}
The weights of the groups are chosen as to approximate the relative area of pixels covered by the respective group in the images (which gives a rough estimate of the relative number of polyps in each group). 

The label $y^{(i)}$ for image $i$ was computed according to  
\begin{equation} \label{eq:definition_labels}
	y^{(i)} = \begin{cases}
		1 &\text{if } \eta_i \geq \frac{1}{2} \\
		0 &\text{otherwise,}
	\end{cases}
\end{equation}
where $y^{(i)} = 1$ corresponds to \emph{extended} and $y^{(i)} = 0$ to \emph{retracted}. 

\begin{figure}
\centering
\includegraphics[width=\linewidth]{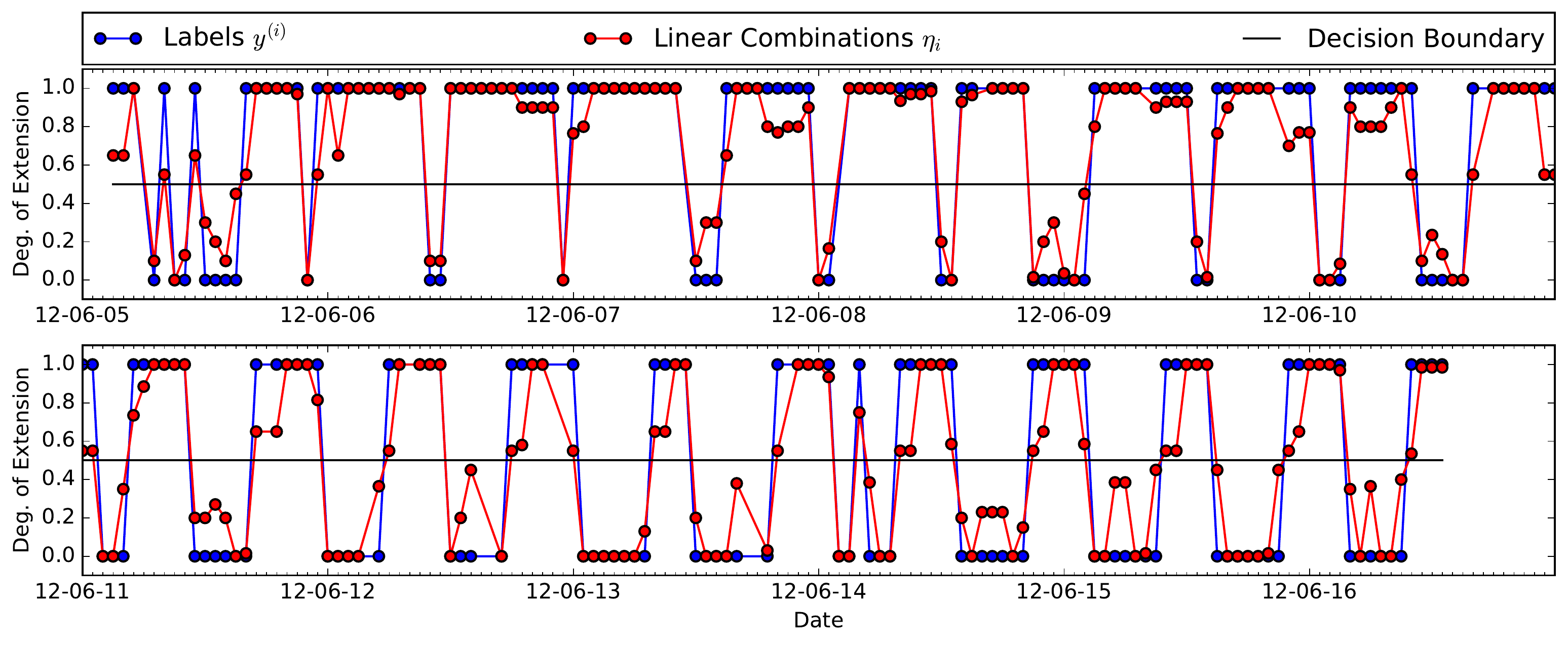}
\caption{Linear combination series of observations $(\eta_i)_{i}$ (red; see \autoref{eq:linear_combination_observations}) and binary extended/retracted label series $(y^{(i)})_{i}$ (blue; see \autoref{eq:definition_labels}).}
\label{fig:observations}
\end{figure}

The resulting series $(\eta_i)_{i}$ and $(y^{(i)})_{i}$ are displayed in \autoref{fig:observations}. 
From the $k=258$ images in total, $k_\tn{ext} = 161$ ($62.4\%$) are classified as \emph{extended} and $k_\tn{retr} = 97$ ($37.6\%$) as \emph{retracted}. 

\subsection{Time Series Data}

The temperature, conductivity, and pressure time series sampled in 10 minute intervals by the main lander MLM were transformed into conservative temperature, absolute salinity and $\sigma_\theta$\hyp{}density series according to TEOS-10 \parencite{mcdougall2011}. 
The directions recorded by the up\hyp{} and downward\hyp{}facing ADCPs (given in degrees) were transformed to Cartesian coordinates on the unit cycle via 
\begin{equation}
	\alpha \mapsto \left(\cos\left(\frac{\pi \alpha}{180}\right), \, \sin\left(\frac{\pi \alpha}{180}\right) \right)^T \text{.}
\end{equation}
This mapping ensures that angles close to $0^\circ$ and close to $360^\circ$ are close to each other with respect to the Euclidean metric. 
Using our web\hyp{}based tool OceanTEA,\footnote{\url{https://github.com/a-johanson/oceantea}} it is possible to explore the time series data analyzed in this study online.\footnote{Follow the link provided at: \url{https://github.com/a-johanson/paragorgia-arborea-activity}} 

We linearly interpolated each of the oceanographic time series at the timestamps $t_i$ of the labeled photographs described in the previous section. 
To obtain stationary time series, we additionally calculated the first\hyp{}order derivative of each series at the timestamps $t_i$. 
However, since the derivative time series turned out to be irrelevant in our subsequent analysis, we ignored them. 
Furthermore, we calculated lagged time series, at first, at $1\text{ h}, 2\text{ h},\ldots,11\text{ h}$ lags covering a full tidal cycle. 
As, in our analysis, only lags between 2 and 4 h proved to be relevant features, we considered only those lags (lags between 2 and 4 h with a finer resolution of 10 minutes also did not improve our results and have, therefore, been ignored). 

\begin{table}
	\centering
	\begin{tabular}{lrr}
		\toprule
		Series & \# of PCs & Variance explained \\ \midrule
		Velocity up & 1 & $91.5\%$ \\
		Cartesian direction up & 1 & $82.9\%$ \\
		Velocity down & 3 & $71.0\%$ \\
		Cartesian direction down & 3 & $67.0\%$ \\ \bottomrule
	\end{tabular}
	\caption{Results of PCA for ADCP time series.}
	\label{tab:pca_adcp}
\end{table}

To compress the ADCP data, we applied principal component analysis (PCA) to the multivariate (multiple bins) series for flow direction up, flow direction down, flow velocity up, and flow velocity down. 
We retained as many principal components (PCs) per time series as necessary to capture at least $75\%$ of the variance but at most $3$ PCs. 
The upward\hyp{}facing ADCP series can be compressed more efficiently via PCA than the data of the downward\hyp{}facing ADCP (see \autoref{tab:pca_adcp}). 

\begin{table}
	\centering
	\begin{tabular}{lccccc}
		\toprule
		Feature & Base Series & Derivative & Lag 2 h & Lag 3 h & Lag 4 h \\ \midrule
		Conservative temperature & \ding{52} & \ding{56} & \ding{52} & \ding{52} & \ding{52} \\
		Absolute salinity & \ding{52} & \ding{56} & \ding{52} & \ding{52} & \ding{52} \\
		$\sigma_\theta$-density & \ding{52} & \ding{56} & \ding{52} & \ding{52} & \ding{52} \\
		PCA direction up (1 PC) & \ding{52} & \ding{56} & \ding{52} & \ding{52} & \ding{52} \\
		PCA velocity up (1 PC) & \ding{52} & \ding{56} & \ding{52} & \ding{52} & \ding{52} \\
		PCA direction down (3 PCs) & \ding{52} & \ding{56} & \ding{52} & \ding{52} & \ding{52} \\
		PCA velocity down (3 PCs) & \ding{52} & \ding{56} & \ding{52} & \ding{52} & \ding{52} \\ \bottomrule
	\end{tabular}
	\caption{Features extracted from the time series data provided by the MoLab main lander. PCA direction and velocity up/down indicate data from the upward\hbox{-/}downward\hyp{}facing ADCP.}
	\label{tab:features}
\end{table}

In total, we obtained the $n=44$ features given in \autoref{tab:features} from the time series data provided by the MoLab main lander. 
The features are arranged in a matrix $(X_{ij})_{i,j} \in \RR^{k\times n}$ so that $X_{ij}$ represents the $j$\hyp{}th feature at time $t_i$. 
We write $x^{(i)}$ to denote the vector $\left(X_{i,1},\ldots,X_{i,n}\right)^T$ of all $n$ features at time $t_i$.

\subsection{Classification Method}

We employed supervised learning techniques \parencite{shalev2014} to learn a classifier that, given a feature vector $x\in \RR^n$, predicts whether the coral polyps are extended or retracted. 
The overall idea of supervised machine learning techniques is to come up with a general mathematical model (i.e., a class of functions---a so\hyp{}called \emph{hypothesis class}) of how some variable depends on certain other variables and to use data to \enquote{learn} the parameters of this general model to make predictions about the value of the dependent variable for so far unobserved states of the independent variables. 
In our case, we want to study if and how coral polyp extension/retraction depends on the oceanographic signals we sampled. 
We assigned a binary \emph{label} $y^{(i)}$ to each photograph of the coral colonies as described in \autoref{eq:definition_labels} in \autoref{subsec:photographs}: 
$y^{(i)} = 1$ represents extended polyps, $y^{(i)} = 0$ retracted polyps. 
Each photograph is associated with certain \emph{features}, such as the temperature at the time the image was taken. 
We group the $n$ features associated with the $i$\hyp{}th image into a feature vector $x^{(i)} \in \RR^n$. 
Now, we select a function class $h_\theta(x)$ parametrized with some parameters $\theta \in \RR^{k}$ that we believe to capture the general relationship between our dependent variable (extension/retraction of the polyps in the photographs) and the features best (e.g., we could assume a linear or quadratic relationship between them). 
Once a certain hypothesis class has been identified, we \enquote{learn} a specific hypothesis from our observations by solving an optimization problem to find values for the parameters $\theta$ that minimize the error between our observations $y^{(i)}$ and the predictions $h_\theta(x^{(i)})$ (i.e., by fitting the hypothesis function to our data). 
This gives us a \emph{classifier} function that predicts a label (retracted/extended) for all possible states of our features (esp.~for so far unobserved states). 
Furthermore, if the hypothesis $h_\theta$ has a form that is readily interpretable, we can quantify the impact each oceanographic signal has on the state of the coral polyps. 

A linear classifier (i.e., a classifier that fits a hyperplane to our feature space and classifies data points on one side of the hyperplane as \enquote{extended} and those on the other side as \enquote{retracted}) suits our task best for two reasons: 
\begin{enumerate}
	\item A linear classifier avoids overfitting to some degree. 
	\item A linear learned hypothesis is readily interpretable with respect to the influence of individual features on the classification process (after all, we are more interested in understanding which features govern the decision than we are in the predictions themselves). 
\end{enumerate}
We chose logistic regression (LR) as our learning algorithm over linear support vector machines (SVM) and decision trees (although decision trees are not linear classifiers, they provide easily interpretable hypotheses). 
LR achieved better prediction accuracy than an SVM and decision trees had to overfit to achieve the accuracy of LR.

\subsubsection{Logistic Regression}

\begin{figure}
	\centering
	\includegraphics[width=0.8\linewidth]{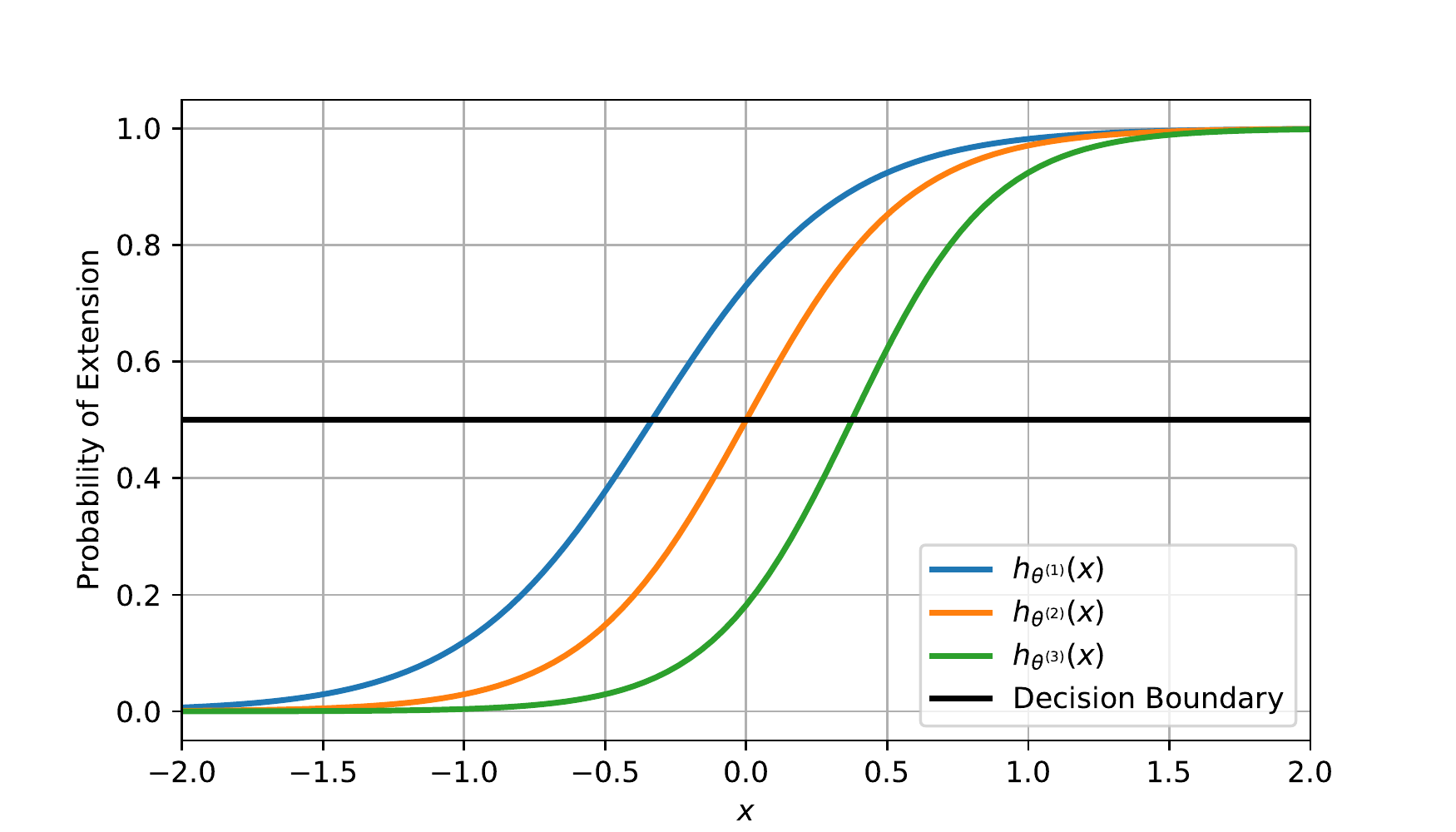}
	\caption{Members of the logistic functions hypothesis class for a one\hyp{}dimensional feature space and different parameter vectors $\theta^{(1)}$, $\theta^{(2)}$, and $\theta^{(3)}$. In these cases, the more positive the value of $x$, the more likely it is that the polyps are extended and vice versa. We classify probabilities greater than or equal to $\frac{1}{2}$ as extended and those smaller than $\frac{1}{2}$ as retracted.}
	\label{fig:lr}
\end{figure}

For LR, we have $m$ training examples $(x^{(i)}, y^{(i)})$ with inputs $x^{(i)} \in \RR^n$ ($n$ features) and labels $y^{(i)} \in \{0,1\}$, $i=1,\ldots,m$. 
The hypothesis class we are learning consists of the logistic functions 
\begin{equation}
	h_\theta \colon \RR^n \to (0,1), \; x \mapsto \frac{1}{1+\exp(-\theta^T x)}
\end{equation}
parametrized by $\theta \in \RR^{n+1}$.
Here, we set $x_0=1$ so that we achieve a convenient notation to include the intercept term $\theta_0$:  
\begin{equation}
	\theta^T x = \theta_0 + \sum_{j=1}^n \theta_j x_j
\end{equation}
Since the range of $h_\theta$ is $(0,1)$ for all $\theta$, we can express our classification problem in a probabilistic framework: 
We set the probability that, given an input $x\in\RR^n$ and parameters $\theta \in \RR^{n+1}$, the coral polyps are extended as 
\begin{equation}
	P(y=1|x;\theta) = h_\theta(x)
\end{equation}
and, therefore, the probability that they are retracted is given by 
\begin{equation}
	P(y=0|x;\theta) = 1-h_\theta(x) \text{.}
\end{equation}
This can be written as 
\begin{equation}
	p(y|x;\theta) = h_\theta(x)^y (1 - h_\theta(x))^{1-y} \text{.}
\end{equation}
To classify a new input vector $x \in \RR^n$, we simply check whether, given $x$, our model predicts the coral polyps to be more likely extended or to be more likely retracted (see \autoref{fig:lr}). 
We use $y_\theta(x)$ to denote the prediction generated by the model parametrized with $\theta$. 

To obtain suitable parameters $\theta$ for our model, we maximize the likelihood $L(\theta)$ for our training examples (which we assume to be independent) by using a gradient method to find 
\begin{equation} \label{eq:max_likelihood}
	\arg\max_\theta \; L(\theta) = \arg\max_\theta \; \prod_{i=1}^{m} p\left(y^{(i)} | x^{(i)} ; \theta\right) \text{.}
\end{equation}
As it is difficult to compute the gradient of the right hand side of \autoref{eq:max_likelihood} with respect to $\theta$, it is more convenient to maximize the log-likelihood via 
\begin{equation}
	\arg\max_\theta \; \log(L(\theta)) = \arg\max_\theta \; \sum_{i=1}^{m} \log\left(p\left(y^{(i)} | x^{(i)} ; \theta\right)\right) \text{,}
\end{equation} 
which yields the same solution by virtue of the monotonicity of the logarithm. 
For solving this optimization problem, we employ LIBLINEAR \parencite{fan2008} and apply $\ell_2$ regularization. 


Before fitting the hypothesis to the data, we centered and scaled each feature independently by subtracting its sample mean and by dividing by its sample standard deviation (computed from the biased sample variance): 
\begin{align}
	\mu_j &= \frac{1}{m} \sum_{i=1}^{m} X_{ij} \label{eq:mean} \\
	s_j &= \sqrt{\frac{1}{m} \sum_{i=1}^{m} (X_{ij} - \mu_j)^2} \label{eq:std_dev} \\
	\hat{X}_{ij} &= \frac{X_{ij} - \mu_j}{s_j} \label{eq:normalization}
\end{align}
This normalization of the input allows us to compare the relative importance of individual features in the model. 

\subsubsection{Assessing Feature Importance}

The absolute value of $\theta_j$, $j=1,\ldots,n$ reflects the importance of the $j$\hyp{}th feature in the model (if $\theta_j=0$, the $j$\hyp{}th feature has no influence on model results at all). 
However $| \theta_j |$, does not \emph{linearly} reflect feature importance. 
To derive a linear measure of importance, one has to analyze changes in the odds per unit of in-/decrease of feature $j$. 

The odds for the polyps to be extended or retracted according to our model are given by: 
\begin{align}
	o_{ext} (x) &= \frac{P(y=1|x;\theta)}{P(y=0|x;\theta)} = \frac{h_\theta(x)}{1-h_\theta(x)} \\
			&= \frac{1}{1+\exp(-\theta^T x)} \cdot \frac{1}{1 - \frac{1}{1+\exp(-\theta^T x)}}\\
			&= \frac{1}{\exp(-\theta^T x)} = \exp(\theta^T x)\\
	o_{retr} (x) &= \frac{1}{o_{ext} (x)} = \exp(-\theta^T x)
\end{align}
The odds multiply by $\exp(\theta_j)$ and $\exp(-\theta_j)$, respectively, for every $1$\hyp{}unit increase in $x_j$: 
\begin{align}
	o_{ext} (x+1\cdot e_j) &= \exp(\theta^T (x+e_j)) \\
		&= \exp(\theta^T x) \cdot \exp(\theta^T e_j) \\
		&= \exp(\theta_j) \cdot o_{ext} (x) \\
	o_{retr} (x+1\cdot e_j) &= \exp(-\theta^T (x+e_j)) \\
		&= \exp(-\theta_j) \cdot o_{retr} (x)
\end{align}

To assess the effect of $\theta_j$ on changes in the odds, we define 
\begin{equation}
	\Delta o_j = 100 \cdot (\exp(|\theta_j|) - 1) \% \text{.}
\end{equation}
$\Delta o_j$ expresses by how many percent the odds change for every 1\hyp{}unit increase in $x_j$ (if $\theta_j \geq 0$, for $o_{ext}$, otherwise, for $o_{retr}$).

\subsubsection{Recursive Feature Elimination} \label{subsec:rfe}

For selecting only a subset of our $n$ features, we apply recursive feature elimination (RFE). 
RFE first trains a model with all features and then iteratively removes the feature with the smallest absolute weight $|\theta_j|$ until only the desired number of features is left.

\subsubsection{Model Validation and Data Sparsity} \label{subsec:validation_sparsity}

Once we obtained a hypothesis function for a given set of features and training data (which we used to optimize the parameters of the hypothesis), we need to assess how well our classifier is able to predict the extension/retraction of coral polyps for feature values not used in the training process (i.e., how well our classifier generalizes to new data). 
To achieve this, one can employ a technique known as \emph{cross validation}, in which we train our classifier only with a subset of the data available to us and test its performance on the rest of the data, which has not been used in the training process. 

To validate our classifiers, we employed cross validation with a partition of our labeled data $(x^{(i)}, y^{(i)})$ into a training set $\mathcal{T} \in \mathcal{P}(\{1,\ldots,m\})$ ($60\%$) and a validation set $\mathcal{V} \in \mathcal{P}(\{1,\ldots,m\})$ ($40\%$), where $\mathcal{P}$ denotes the power set. 
Note that we determined $\mu_j$ (\autoref{eq:mean}) and $s_j$ (\autoref{eq:std_dev}) for the training set $\mathcal{T}$ only and applied the normalization operation given in \autoref{eq:normalization} to both the training and the validation set using these values. 

The accuracy of a model with parameters $\theta \in \RR^{n+1}$ on the validation set is defined by 
\begin{equation}
	a(\theta, \mathcal{V}) = 1 - \frac{1}{|\mathcal{V}|} \sum_{i\in \mathcal{V}} \left| y^{(i)} - y_\theta(x^{(i)}) \right| \text{.}
\end{equation}

To assess whether adding more features to a model makes its predictions significantly better, we employ the likelihood\hyp{}ratio test \parencite{casella2001}. 
Its test statistic is computed by 
\begin{equation}
	\rho = 2 \left( \log(L(\theta^{(2)})) - \log(L(\theta^{(1)})) \right) \text{,}
\end{equation}
which compares the log likelihood of the parameters of the more complex model ($\theta^{(2)}$) with the log likelihood of the ones of the less complex model ($\theta^{(1)}$). 
It holds $\rho \sim \chi^2$ with degrees of freedom equal to the number of parameters that are constrained in the lower\hyp{}complexity model in comparison to the other. 
This implies that when adding a single feature to a model, it only provides a significant improvement with $95\%$ confidence if $\rho > 3.841$. 

Since our data set it relatively small ($k=258$ images), different random partitions of the data into training and validation set lead to different models being chosen during model training (especially due to differing outcomes in the RFE process). 
To make our approach more stable with respect to different partitions in training and validation set, we repeat each model training (including RFE) $\iota=1000$ times and average all relevant statistics (such as validation accuracy $a$ and feature weights $\theta$). 
We found that this approach converges for iteration counts as low as $\iota=100$. 

All analyses reported on in this paper were implemented using Python 3.5.1\footnote{\url{http://www.python.org}} with scikit-learn 0.17.1.\footnote{\url{http://scikit-learn.org}} 
Our implementation is available online.\footnote{\url{https://github.com/a-johanson/paragorgia-arborea-activity}}

\section{Results}

\subsection{Optimal Number of Features}

\begin{table}
	\centering \small
	\begin{tabular}{lrrrr}
		\toprule
		\# Feat. & Avg. Accuracy $a$ & Avg. Likelihood & Log Likelihood & Likelihood Ratio $\rho$ \\ \midrule
		   1   &  80.243\%  & 4.62161e-10  &  -21.495107   &   --- \\
		   2   &  83.682\%  & 1.75329e-08  &  -17.859185   &   \textbf{7.271844} \\
		   3   &  85.537\%  & 8.83761e-08  &  -16.241664   &   3.235043 \\
		   4   &  86.259\%  & 5.54831e-07  &  -14.404603   &   3.674122 \\
		   5   &  86.866\%  & 1.34843e-06  &  -13.516569   &   1.776068 \\
		   6   &  \textbf{87.086\%}  & 3.02838e-06  &  -12.707484   &   1.618169 \\
		   7   &  87.038\%  & 6.23238e-06  &  -11.985752   &   1.443465 \\
		   8   &  86.933\%  & 9.26830e-06  &  -11.588910   &   0.793683 \\
		   9   &  86.839\%  & 1.26937e-05  &  -11.274407   &   0.629006 \\
		  10   &  86.676\%  & 1.57651e-05  &  -11.057714   &   0.433387 \\
		  11   &  86.528\%  & 1.61683e-05  &  -11.032457   &   0.050513 \\
		  12   &  86.377\%  & 1.89536e-05  &  -10.873515   &   0.317885 \\
		  13   &  86.303\%  & 2.21165e-05  &  -10.719185   &   0.308660 \\
		  14   &  86.171\%  & 2.53495e-05  &  -10.582751   &   0.272868 \\
		  15   &  86.146\%  & 2.94565e-05  &  -10.432597   &   0.300306 \\
		  16   &  86.098\%  & 3.24890e-05  &  -10.334608   &   0.195979 \\
		  17   &  86.040\%  & 3.61036e-05  &  -10.229117   &   0.210982 \\
		  18   &  85.990\%  & 3.84051e-05  &  -10.167321   &   0.123592 \\
		  19   &  85.978\%  & 4.22247e-05  &  -10.072505   &   0.189633 \\
		  20   &  85.923\%  & 4.66312e-05  &   -9.973240   &   0.198529 \\
		  21   &  85.930\%  & 4.95968e-05  &   -9.911584   &   0.123312 \\
		  22   &  85.981\%  & 5.31794e-05  &   -9.841839   &   0.139491 \\
		  23   &  85.941\%  & 5.68688e-05  &   -9.774764   &   0.134149 \\
		  24   &  85.920\%  & 6.05087e-05  &   -9.712724   &   0.124081 \\
		  25   &  85.927\%  & 6.18050e-05  &   -9.691526   &   0.042395 \\
		  26   &  85.940\%  & 6.20666e-05  &   -9.687303   &   0.008446 \\
		  27   &  85.945\%  & 6.16865e-05  &   -9.693446   &  -0.012286 \\
		  28   &  85.980\%  & 6.39735e-05  &   -9.657042   &   0.072808 \\
		  29   &  85.965\%  & 6.51849e-05  &   -9.638283   &   0.037518 \\
		  30   &  85.963\%  & 6.73129e-05  &   -9.606159   &   0.064247 \\
		  31   &  85.937\%  & 6.85712e-05  &   -9.587638   &   0.037042 \\
		  32   &  85.962\%  & 6.81004e-05  &   -9.594527   &  -0.013778 \\
		  33   &  85.954\%  & 6.86092e-05  &   -9.587084   &   0.014887 \\
		  34   &  85.938\%  & 6.89208e-05  &   -9.582552   &   0.009064 \\
		  35   &  85.952\%  & 7.04471e-05  &   -9.560648   &   0.043807 \\
		  36   &  85.961\%  & 7.03325e-05  &   -9.562277   &  -0.003258 \\
		  37   &  85.941\%  & 7.10724e-05  &   -9.551811   &   0.020932 \\
		  38   &  85.941\%  & 7.22947e-05  &   -9.534760   &   0.034103 \\
		  39   &  85.926\%  & 7.24531e-05  &   -9.532571   &   0.004376 \\
		  40   &  85.942\%  & 7.36322e-05  &   -9.516428   &   0.032286 \\
		  41   &  85.927\%  & 7.32244e-05  &   -9.521982   &  -0.011108 \\
		  42   &  85.930\%  & 7.32091e-05  &   -9.522191   &  -0.000417 \\
		  43   &  85.935\%  & 7.32887e-05  &   -9.521105   &   0.002172 \\
		  44   &  85.932\%  & 7.34663e-05  &   -9.518684   &   0.004841 \\
		\bottomrule
	\end{tabular}
	\caption{Average validation accuracy and likelihood\hyp{}ratio test statistic for adding a single feature depending on the number of features in the model. Optimal/significant values printed in bold ($\alpha=0.05$).}
	\label{tab:optimal_num_features}
\end{table}

\autoref{tab:optimal_num_features} displays the results of training models with $n=1$ to $44$ features (selected by RFE as described in \autoref{subsec:rfe}) and averaging the validation accuracy and parameter likelihood for each $n$ over $\iota=1000$ iterations (\autoref{subsec:validation_sparsity}). 
The likelihood ratio shown in row $i$ of this table is calculated using the log likelihood of rows $i$ and $i-1$ (constraining one feature). 

While optimal validation accuracy is achieved for $n=6$ features, only the first likelihood ratio is significant for $\alpha=0.05$. 
Therefore, continuing to add a single feature at a time to a model with $n=2$ features does not significantly increase prediction accuracy. 
However, adding four features to a two feature model (to obtain $n=6$ features, which results in optimal validation accuracy), does significantly increase prediction performance ($\rho=2\cdot(-12.707484+17.859185) \approx 10.303402 > 9.488$). 

In the following, we take a closer look at models with $n=2$ and $n=6$ features to analyze which of the features---i.e., oceanographic signals---has the greatest impact on coral polyp behavior. 
The model with $n=2$ features is distinguished from the others by being the model with the smallest number of features to which adding a single feature does not further improve prediction accuracy significantly. 
The $n=6$ feature model stands out because it achieves the highest overall predication accuracy.

\subsection{Most Informative Features}

\begin{table}
	\centering \small
	\begin{tabular}{lrrrr}
		\toprule
		Feature                    & Times Chosen & Avg. $\theta_i$ & Avg. $| \theta_i |$ & $\Delta o_i$ \\ \midrule
		Direction up, 4h lag, PC1  &       63.6\% &        -1.77981 &             1.77981 &     +492.9\% \\
		Direction up, 3h lag, PC1  &       48.7\% &        -1.74120 &             1.74120 &     +470.4\% \\
		Velocity up, 2h lag, PC1   &       42.9\% &        -1.23761 &             1.23761 &     +244.7\% \\
		Velocity down, 3h lag, PC3 &       14.2\% &         1.02258 &             1.02258 &     +178.0\% \\
		Velocity down, PC1         &       13.9\% &        -1.26549 &             1.26549 &     +254.5\% \\
		Direction up, PC1          &        8.5\% &         1.00027 &             1.00027 &     +171.9\% \\
		Velocity down, 2h lag, PC2 &        3.1\% &         0.91429 &             0.91429 &     +149.5\% \\
		Direction down, PC1        &        1.6\% &        -0.90477 &             0.90477 &     +147.1\% \\
		Velocity down, 3h lag, PC2 &        1.1\% &         0.94943 &             0.94943 &     +158.4\% \\ \bottomrule
	\end{tabular}
	\caption{Average feature occurrence in a model with $n=2$ features (displaying only features that are chosen in more than $1\%$ of the iterations).}
	\label{tab:2feature_model}
\end{table}

\begin{table}
	\centering \small
	\begin{tabular}{lrrrr}
		\toprule
		Feature                     & Times Chosen & Avg. $\theta_i$ & Avg. $| \theta_i |$ & $\Delta o_i$ \\ \midrule
		Direction up, 3h lag, PC1   &       97.3\% &        -1.43840 &             1.43840 &     +321.4\% \\
		Velocity up, 2h lag, PC1    &       84.8\% &        -1.02575 &             1.02575 &     +178.9\% \\
		Direction up, 4h lag, PC1   &       80.8\% &        -1.31124 &             1.31124 &     +271.1\% \\
		Velocity down, 3h lag, PC2  &       74.0\% &         0.86593 &             0.86593 &     +137.7\% \\
		Velocity down, 3h lag, PC3  &       64.3\% &         0.93843 &             0.93843 &     +155.6\% \\
		Velocity down, 2h lag, PC2  &       60.9\% &         0.86914 &             0.86914 &     +138.5\% \\
		Direction up, PC1           &       25.7\% &         0.93158 &             0.93158 &     +153.9\% \\
		Velocity down, PC1          &       23.8\% &        -0.92486 &             0.92486 &     +152.2\% \\
		Velocity down, 4h lag, PC3  &       15.6\% &         0.76278 &             0.76278 &     +114.4\% \\
		Direction down, 3h lag, PC2 &       10.5\% &        -0.77474 &             0.77474 &     +117.0\% \\
		Direction down, PC1         &        7.8\% &        -0.90753 &             0.90753 &     +147.8\% \\
		Velocity down, PC3          &        5.7\% &        -0.76523 &             0.76523 &     +114.9\% \\
		Temperature (cons.), 3h lag &        5.6\% &        -0.69882 &             0.69882 &     +101.1\% \\
		Velocity down, 2h lag, PC3  &        5.3\% &         0.81318 &             0.81318 &     +125.5\% \\ \bottomrule
	\end{tabular}
	\caption{Average feature occurrence in a model with $n=6$ features (displaying only features that are chosen in more than $5\%$ of the iterations).}
	\label{tab:6feature_model}
\end{table}

To examine which features are most informative with respect to whether the coral polyps are extended or retracted, we list in Tables \ref{tab:2feature_model} and \ref{tab:6feature_model} for models with $n=2$ and $n=6$ features which features are chosen most often by RFE and how sensitive the models are to each feature on average ($\Delta o_i$).
Note that how often feature $i$ is chosen is highly correlated with $\Delta o_i$. 

We can observe that the most important features for the classifiers are all related to the measurements of currents by the upward\hyp{} and downward facing ADCPs. 
Among the features not directly representing current measurements, only conservative temperature at a lag of 3 h is chosen in $5.6\%$ of the models with $n=6$ features. 

Among the features derived from the ADPC data, those at lags of 2 to 4 h play the most important role. 
Note that the features from the upward\hyp{}facing ADCP have greater influence on predictions than those obtained from the downward\hyp{}facing instrument. 
Also note that for the downward\hyp{}facing features, the second and third principal component are often more important than the first. 

\begin{table}
	\centering \small
	\begin{tabular}{lrrrr}
		\toprule
		Feature                         & Avg. Accuracy $a$ & Avg. $\theta_1$ & Avg. $| \theta_1 |$ & $\Delta o_i$ \\ \midrule
		Direction up, 3h lag, PC1       &           \textbf{80.83\%} &        -1.60309 &             1.60309 &      396.8\% \\
		Direction up, 4h lag, PC1       &           \textbf{80.49\%} &        -1.72458 &             1.72458 &      461.0\% \\
		Direction down, 4h lag, PC1     &           \textbf{76.71\%} &         1.24388 &             1.24388 &      246.9\% \\
		Velocity down, 3h lag, PC1      &           \textbf{75.14\%} &        -1.24781 &             1.24781 &      248.3\% \\
		Direction down, 3h lag, PC2     &           \textbf{74.98\%} &        -0.77129 &             0.77129 &      116.3\% \\
		Velocity down, 2h lag, PC1      &           \textbf{72.54\%} &        -1.08347 &             1.08347 &      195.5\% \\
		Velocity down, 4h lag, PC1      &           \textbf{72.09\%} &        -0.83732 &             0.83732 &      131.0\% \\
		Direction down, 3h lag, PC1     &           \textbf{68.53\%} &         0.90314 &             0.90314 &      146.7\% \\
		Direction up, 2h lag, PC1       &           \textbf{66.32\%} &        -0.70869 &             0.70869 &      103.1\% \\
		Velocity up, 2h lag, PC1        &           \textbf{66.27\%} &        -0.92686 &             0.92686 &      152.7\% \\
		Direction down, 2h lag, PC2     &           \textbf{65.85\%} &        -0.50002 &             0.50002 &       64.9\% \\
		Direction down, 4h lag, PC2     &           \textbf{65.57\%} &        -0.42829 &             0.42829 &       53.5\% \\
		Velocity up, 3h lag, PC1        &           \textbf{65.50\%} &        -0.78162 &             0.78162 &      118.5\% \\
		Direction down, PC3             &           \textbf{64.43\%} &        -0.59813 &             0.59813 &       81.9\% \\
		Direction up, PC1               &           \textbf{63.51\%} &         0.48897 &             0.48897 &       63.1\% \\
		Velocity down, PC3              &           \textbf{63.04\%} &        -0.33249 &             0.33262 &       39.4\% \\
		Velocity down, 3h lag, PC3      &           \textbf{62.88\%} &         0.46583 &             0.46583 &       59.3\% \\
		Velocity down, 4h lag, PC3      &           62.51\% &         0.13308 &             0.14454 &       14.2\% \\
		Salinity (abs.), 3h lag         &           62.47\% &        -0.01264 &             0.07779 &        1.3\% \\
		$\sigma_\theta$-density, 3h lag &           62.47\% &        -0.00811 &             0.07768 &        0.8\% \\
		Velocity up, PC1                &           62.46\% &         0.00827 &             0.08383 &        0.8\% \\
		Direction down, PC2             &           62.43\% &        -0.02661 &             0.08256 &        2.7\% \\
		Velocity down, PC1              &           62.43\% &         0.02170 &             0.08302 &        2.2\% \\
		$\sigma_\theta$-density, 4h lag &           62.38\% &         0.04302 &             0.08913 &        4.4\% \\
		Temperature (cons.), 2h lag     &           62.33\% &        -0.03894 &             0.08734 &        4.0\% \\
		Salinity (abs.), 4h lag         &           62.33\% &         0.05431 &             0.09293 &        5.6\% \\
		Velocity down, 4h lag, PC2      &           62.31\% &        -0.08802 &             0.11215 &        9.2\% \\
		Temperature (cons.), 3h lag     &           62.30\% &        -0.06065 &             0.09312 &        6.3\% \\
		Salinity (abs.)                 &           62.28\% &        -0.11244 &             0.12566 &       11.9\% \\
		Temperature (cons.)             &           62.26\% &        -0.05773 &             0.09588 &        5.9\% \\
		Salinity (abs.), 2h lag         &           62.26\% &        -0.12469 &             0.13666 &       13.3\% \\
		$\sigma_\theta$-density, 2h lag &           62.23\% &        -0.12566 &             0.13745 &       13.4\% \\
		$\sigma_\theta$-density         &           62.22\% &        -0.11133 &             0.12500 &       11.8\% \\
		Velocity down, 3h lag, PC2      &           62.16\% &         0.35918 &             0.35918 &       43.2\% \\
		Velocity down, 2h lag, PC3      &           62.02\% &         0.37153 &             0.37153 &       45.0\% \\
		Direction down, PC1             &           61.87\% &        -0.41082 &             0.41082 &       50.8\% \\
		Velocity down, PC2              &           61.59\% &         0.23752 &             0.23855 &       26.8\% \\
		Temperature (cons.), 4h lag     &           61.39\% &         0.16489 &             0.16938 &       17.9\% \\
		Velocity down, 2h lag, PC2      &           61.19\% &         0.37499 &             0.37499 &       45.5\% \\
		Velocity up, 4h lag, PC1        &           60.57\% &        -0.46802 &             0.46802 &       59.7\% \\
		Direction down, 2h lag, PC3     &           59.98\% &        -0.30947 &             0.30965 &       36.3\% \\
		Direction down, 2h lag, PC1     &           59.82\% &         0.46463 &             0.46463 &       59.1\% \\
		Direction down, 4h lag, PC3     &           59.61\% &        -0.21919 &             0.22020 &       24.5\% \\
		Direction down, 3h lag, PC3     &           59.56\% &        -0.25992 &             0.26024 &       29.7\% \\ \bottomrule
	\end{tabular}
	\caption{Average data of models using only one feature ($n=1$). Average accuracies significantly above $\frac{k_\tn{ext}}{k} \approx 62.4\%$ are printed in bold ($\alpha=0.05$).}
	\label{tab:all_1feature_models}
\end{table}

Since, according to \autoref{tab:optimal_num_features}, the accuracy of classifiers using only a single feature is not considerably worse than that of classifiers using two or six, we analyzed how well each feature is able to predict the degree of polyp extension on its own. 
For each of our $44$ features in total, we trained $\iota=1000$ models to obtain the averaged data displayed in \autoref{tab:all_1feature_models}. 
These results largely agree with what we found by studying models with two and six features. 
Specifically, the flow direction of the upward\hyp{}facing ADCP with a 3 h lag is the single best predictor of the degree of extension of the coral polyps with an average accuracy of $80.83\%$. 
This feature is closely followed by the flow direction above the main lander system with a 4 h lag ($80.49\%$ avg.\@ accuracy). 

Note that a trivial model which always predicts the polyps to be extended achieves an accuracy of $\frac{k_\tn{ext}}{k} \approx 62.4\%$ (cf.\@ \autoref{subsec:photographs}). 
Therefore, any model that does not attain an accuracy significantly above $62.4\%$ does not actually contain any information about the degree of polyp extension. 

\subsection{Predicted Polyp Behavior}

\begin{table}
	\centering \small
	\begin{tabular}{lp{5cm}r}
		\toprule
		\# Features & Features & Overall Accuracy \\ \midrule
		1   & Direction up, 3h lag, PC1 & 81.8\% \\ \midrule
		2   & Direction up, 4h lag, PC1 \newline Velocity up, 2h lag, PC1 & 86.4\% \\ \midrule
		6   & Direction up, 3h lag, PC1 \newline Velocity up, 2h lag, PC1 \newline Direction up, 4h lag, PC1 \newline Velocity down, 3h lag, PC2 \newline Velocity down, 3h lag, PC3 \newline Velocity down, 2h lag, PC2 & 91.1\% \\
	\bottomrule
	\end{tabular}
	\caption{Features and overall accuracy of the best one-, two- and six\hyp{}feature models.}
	\label{tab:prediction_models}
\end{table}

We trained models using the most\hyp{}chosen features for models with one, two, and six features (see Tables \ref{tab:2feature_model}, \ref{tab:6feature_model}, and \ref{tab:all_1feature_models}). 
A summary of the three models is given in \autoref{tab:prediction_models}. 
For the second feature of the two\hyp{}feature model, we selected \emph{Velocity up, 2h lag, PC1} instead of \emph{Direction up, 3h lag, PC1} because a model with only the direction up series with 3 and 4 h lags does not perform significantly better than the one\hyp{}feature model with only \emph{Direction up, 3h lag, PC1}. 

\begin{figure}
\centering
\includegraphics[width=\linewidth]{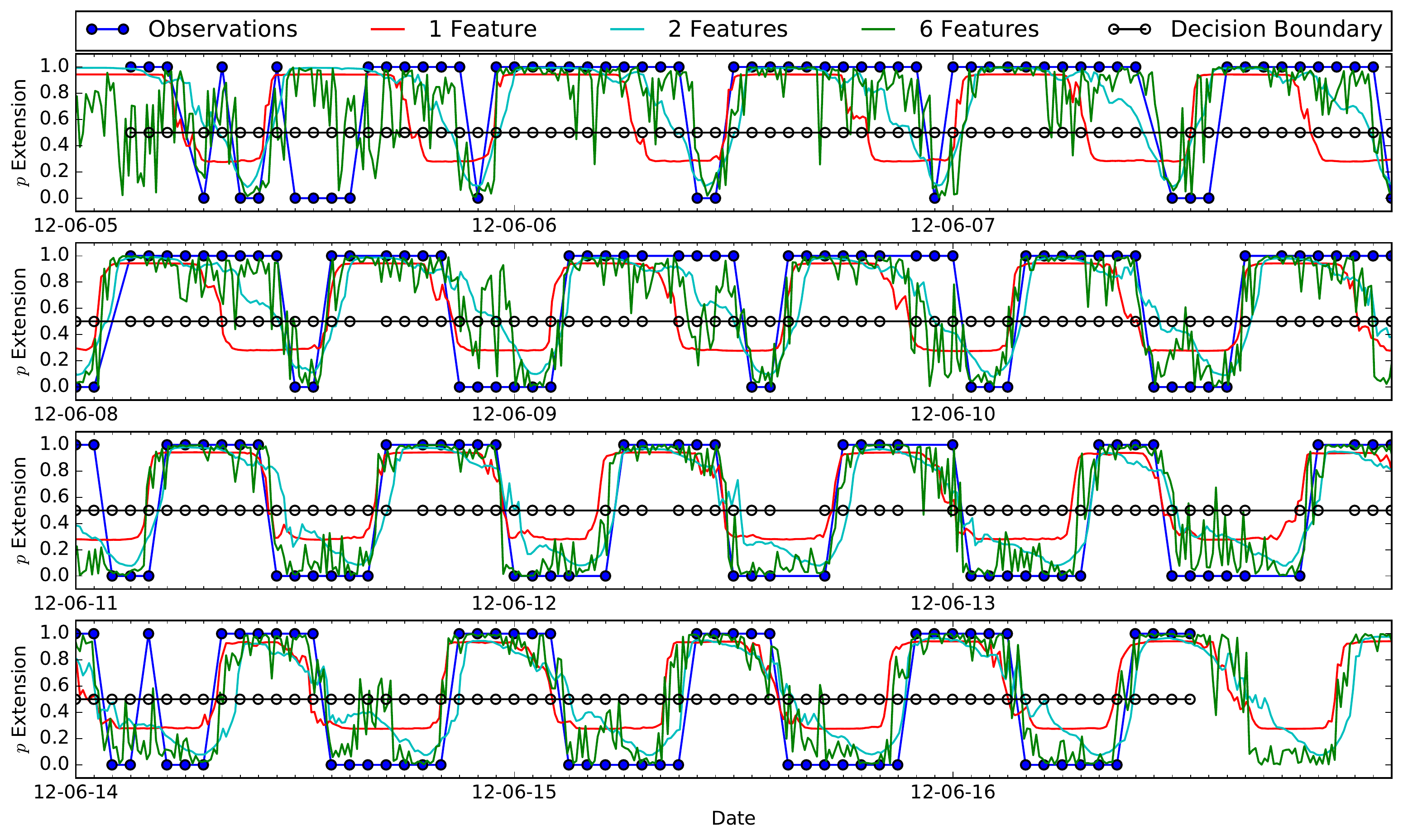}
\caption{Degree of extension of coral polyps according to observations as well as to models with one, two, and six features. For an interactive illustration of this figure follow the link provided at: \url{https://github.com/a-johanson/paragorgia-arborea-activity}}
\label{fig:predictions}
\end{figure}

The predictions of the models in comparison to the observations are shown in \autoref{fig:predictions}. 
An interactive illustration of this figure in which the user can add models using an arbitrary combination of features can be explored online.\footnote{Follow the link provided at: \url{https://github.com/a-johanson/paragorgia-arborea-activity}} 
The observed polyp behavior exhibits clear cycles of an extended state followed by a retracted state with a period of about 12 h. 
Even the one\hyp{}feature model is able to forecast the onset of the extended/retracted cycles of the polyps accurately. 
In the first few extended/retracted cycles, however, the one\hyp{}feature model predicts that the polyps retract sooner as they actually do according to the observations. 
The two\hyp{}feature model performs better at modeling the ending of an extended/retracted cycle but sometimes predicts the onset of the cycle as later than it is in reality. 
While the six\hyp{}feature model is superior to the other two with regard to accuracy at the observations, its predictions are less stable. 
This leads, for example, to the model forecasting the polyps to be extended for quite long time spans during the retracted cycle at the end of June 8 (high overall accuracy is still achieved because at the times of the observations, the correct values are assumed). 
In most cases however, the high sensitivity of the six\hyp{}feature model does not severely affect its ability to accurately predict the sequence of the extended/retracted cycles even at time points between observations.

\subsection{Observed Currents}

\begin{figure}
\centering
\includegraphics[width=0.95\linewidth]{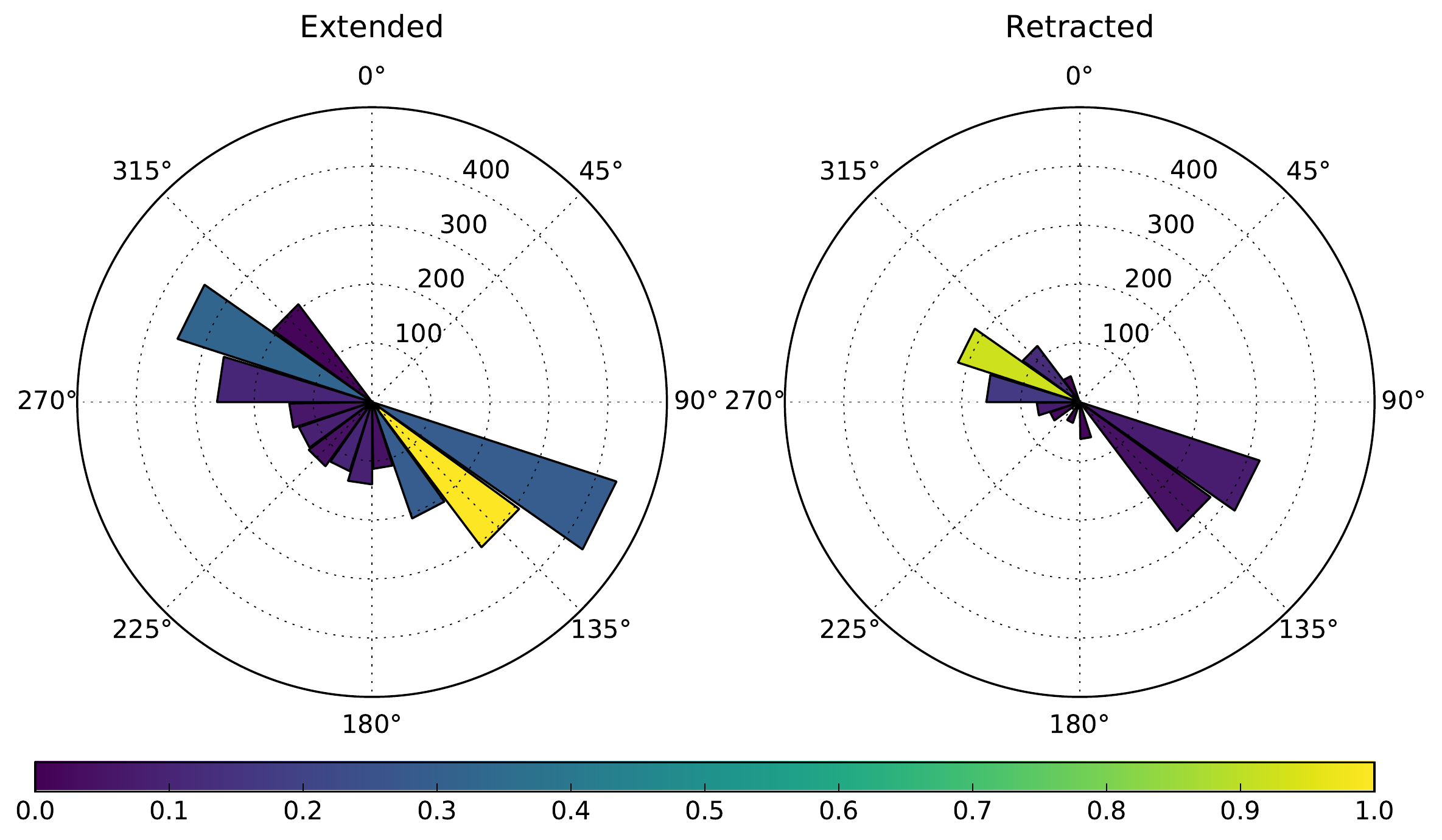}
\caption{Currents observed by the upward\hyp{}facing ADCP for extended and retracted polyps with a lag of 3 h. The length of each segment in the plot represents average velocity in mm s$^{-1}$ and the color represents frequency relative to the number of observations in the most frequent bin (extended, 135$^{\circ}$). Current data is averaged over all analyzed ADCP bins.}
\label{fig:observed_currents}
\end{figure}

The best one\hyp{}feature model uses only the first PC of the direction of the upward\hyp{}facing ADCP with a lag of 3 h. 
From the coefficients of this PC one can compute the average direction of the flow that is associated with extended and with retracted polyps (considering the sign of $\theta_1$). 
According to this, the polyps are most likely to be extended if the direction of the currents above the main lander was $123^\circ$ three hours ago (facing south\hyp{}east; $0^\circ$ corresponds to north and positive angles turn clockwise). 
Accordingly, the polyps are most likely to be retracted with lagged currents flowing in the direction of $303^\circ$ (north\hyp{}west). 
Plotting the properties of the currents observed above the main lander system when the polyps are extended and when they are retracted (\autoref{fig:observed_currents}) supports these results.

\section{Discussion}

\subsection{Tidal Control}

The high prediction accuracy ($>80\%$) achieved by the best one-, two- and six\hyp{}feature models---which all rely only on features derived from the ADCP data---indicates that current regimes play an important role in explaining the behavior of \textit{P. arborea} polyps in the Stjernsund. 
In comparison to the current data, the other oceanographic time series have only a comparably low influence on whether the polyps are extended or retracted. 
In particular, all models derived from a single feature \emph{not} directly associated with the currents are unsuitable for predicting polyp extension: 
the best of such a model uses the absolute salinity with a 3 h lag and only achieves an average accuracy, which is not significantly higher than that of the trivial model predicting the polyps to always be extended (cf.\@ \autoref{tab:all_1feature_models}). 


\subsection{Large\hyp{}Scale vs.\@ Fine\hyp{}Scale Features} \label{subsec:large_vs_fine_scale}

The strong tidal control in the Stjernsund makes it plausible that current patterns govern the behavior of the \textit{P. arborea} corals in our data set to a large extent. 
However, it is noteworthy that the best signal for predicting their polyp behavior is not derived from flow measurements of the water that directly comes into contact with the corals but of the water column further \emph{above} the colonies. 

The large\hyp{}scale pattern of the tidal cycle, which---if represented well in the data---would affect all depth bins in the ADCP measurements equally, can only be captured by a PC with vector components of equal signs for each bin. 
The latter criterion is true for the first PCs of all ADCP time series; for example, it holds: 
\begin{equation}
PC^{(1)}_{\tn{Vel.~up, 3h lag}}= \begin{pmatrix}
	-0.082\\ -0.121\\ -0.124\\ -0.128\\   -0.131 \\
	\vdots\\
	-0.151\\ -0.148\\ -0.147\\ -0.143\\ -0.139
\end{pmatrix} \;\text{ and }\;
PC^{(1)}_{\tn{Vel.~down, 3h lag}}= \begin{pmatrix}
	-0.173\\ -0.117\\ -0.478\\ -0.461\\ -0.416\\ -0.330\\ -0.199\\ -0.181\\ -0.186\\ -0.229\\ -0.271
\end{pmatrix}
\end{equation}
The flows measured by the downward\hyp{}facing ADCP exhibit higher turbulence than those measured by the upward\hyp{}facing (more noise is recorded within each depth bin and the different bins are less correlated, as is evident from the comparably low amount of variance retained by the first components of the PCA). 
Therefore, the large\hyp{}scale pattern of the tidal cycle is less visible in the data obtained from the downward\hyp{}looking instrument and cannot be sampled reliably by the first PC of the respective series. 
This, in turn, reduces the predictive power of the first PCs that are derived from measurements at finer spatial scales (downward\hyp{}facing ADCP) in comparison to the first PCs of the measurements at larger spatial scales (upward\hyp{}facing ADCP). 

In the best six\hyp{}feature model, however, the second and third PCs of the lagged velocity of the flow below the main lander system are included and significantly improve the accuracy of this model (\autoref{tab:prediction_models}). 
These PCs include vector components of varying sign; e.g., it holds that 
\begin{equation}
PC^{(2)}_{\tn{Vel.~down, 2h lag}}= \begin{pmatrix}
	-0.808\\ -0.394\\ -0.118\\ -0.010\\  0.025\\  0.092\\
	 0.169\\  0.174\\   0.157\\   0.182\\  0.227
\end{pmatrix} \;\text{ and }\;
PC^{(3)}_{\tn{Vel.~down, 3h lag}}= \begin{pmatrix}
	0.470\\  -0.336\\ -0.552\\ -0.220\\  0.085\\  0.098\\ 0.191\\ 0.274\\  0.281\\  0.222\\  0.237
\end{pmatrix}
\text{.}
\end{equation}
The mixed signs allow these PCs to represent turbulent events in flow direction and velocity, which are associated especially with changes in water flow direction. 
Therefore, the second and third PCs of the data of the downward\hyp{}facing ADCP can improve the detection of phases in the tidal cycle but at the same time make a model relying on them more prone to reacting to short, random disturbances in the flow to which the corals we observed do not react (explaining the decreased stability of the six\hyp{}feature model). 

It can be concluded that the behavior of the \textit{P. arborea} polyps in the Stjernsund is mostly governed by the large\hyp{}scale patterns of the tidal cycle, which can best be sampled from features at larger spatial scales (i.e., the water above the corals). 
The regular cycles of extended and retracted polyps we observed (\autoref{fig:observations}) would be highly unlikely if smaller\hyp{}scale features played a major role in the behavior of the CWCs. 
In this case, short turbulent events would make the polyps retract and extend their polyps in much faster succession---as it is predicted by the six\hyp{}feature model between many observations. 
Therefore, we have to dismiss the six\hyp{}feature model in favor of the simpler models, which rely solely on said large\hyp{}scale features.


\subsection{Polyp Behavior}

Having established that the extended/retracted cycle of \textit{P. arborea} in the Stjernsund is governed mostly by the tidal cycle, we can now interpret this behavior. 
Since a model featuring only the single\hyp{}most important parameter (the direction of the water flow in the water column above the corals with a lag of three hours) almost has the same predictive power as our best two\hyp{}feature model, we limit our discussion to the one\hyp{}feature model. 

In the following, we describe a full tidal cycle: 
Three hours after the water has begun flowing into the Stjernsund (south\hyp{}eastwards), the coral polyps extend in order to feed on the suspended material that is transported into the sound with the currents. 
They continue to stay extended even after the flow direction reverses and water starts flowing out of the sound (north\hyp{}westwards) for at least three hours (sometimes longer, which cannot be predicted from the feature \emph{Direction up, 3h lag, PC1} and, therefore, limits the accuracy of even the best one\hyp{}feature model; cf.\@ Figures \ref{fig:predictions} and \ref{fig:observed_currents}). 
In this way, the corals are able to continue feeding on food particles that were streamed into the sound with the high tide and are now flowing back out of the sound again. 
When food availability falls below a certain threshold, the polyps retract again and the cycle starts anew.


\section{Conclusions and Outlook}

In this study, we analyzed \textit{in situ} behavior of the cold\hyp{}water coral \textit{Paragorgia arborea} in the Stjernsund using supervised learning techniques. 
We found that the degree of extension of the coral polyps is governed by the current patterns that emerge from the tidal cycle. 
Interestingly, the degree of polyp extension could be predicted more reliably by sampling the laminar water flows above the measurement site than by sampling the more turbulent flows in the direct vicinity of the corals. 
It appears that \textit{P. arborea} in the Stjernsund does not react to shorter changes in the ambient current regime but instead adjusts its behavior in accordance with the large\hyp{}scale pattern of the tidal cycle itself in order to optimize nutrient uptake. 

In the future, it would be worthwhile to include measurements of turbidity and fluorescence in our analysis (which we could not do because of defect instruments). 
It is possible that these time series offer even greater predictive capabilities than the tidally controlled current signal because the quantities they represent affect the corals more immediately (food supply). 
However, when we included turbidity and fluorescence data obtained from the nearby MoLab satellite lander SLM1 into our analysis (\autoref{fig:map_landers}), both features did not exhibit good predictive power, probably because of high noise levels in these series. 
This confirms, once again, that polyp behavior in \textit{P. arborea} in the Stjernsund is governed by large\hyp{}scale instead of small\hyp{}scale features.

\section*{Acknowledgments}

This project was funded by the Cluster of Excellence~80 \enquote{The Future Ocean.} 
The \enquote{Future Ocean} is funded within the framework of the Excellence Initiative by the Deutsche Forschungsgemeinschaft (DFG) on behalf of the German federal and state governments.

We gratefully acknowledge funding for MoLab provided by the Federal Ministry of Education and Research (BMBF) under grant 03F06241. 
The Poseidon cruises POS434 and POS438 were supported by GEOMAR and industry funding (grant A2300414 to P.L.). 
We further thank Norwegian authorities for repeated support and research permissions. 
Last but not least, we would like to thank Andrew Davies for his excellent suggestions in the review process, which greatly improved the paper.

\printbibliography

\end{document}